\documentclass{kapproc} 

\usepackage{procps} 

\usepackage[dvips]{graphicx}

\upperandlowercase

\kluwerbib 

\newcommand*{\Msun}{\ensuremath{\mathrm{M_\odot}}}%
\newcommand*{\Mpc}{\ensuremath{\mathrm{Mpc}}}%
\begin{document}

\articletitle{Specific Star Formation Rates}


\author{A. E. Bauer*,  N. Drory, and  G. J. Hill}

\affil{Department of Astronomy, University of Texas at Austin
1 University Station, \\Austin, Texas 78712}

\email{* amanda@astro.as.utexas.edu}

\begin{abstract}
We present results from a study to determine how star formation
contributes to galaxy growth since redshift $z=1.5$. Using galaxies
from the MUnich Near-Infrared Cluster Survey (MUNICS) and the FORS
Deep Field (FDF), we investigate the specific star formation rate
(SSFR, star formation rate [SFR] per unit galaxy stellar mass) as a
function of galaxy stellar mass and redshift. We test the
compatibility of our results with a sample drawn from a larger volume
using the Sloan Digital Sky Survey.  We find that the SSFR decreases
as galaxy stellar mass increases, suggesting that star formation
contributes more to the growth of low-mass galaxies than high-mass
galaxies at all redshifts in this study. We also find a ridge in the
SSFR that runs parallel to lines of constant SFR and decreases by a
factor of 10 from $z=1$ to today, matching the results of the
evolution in SFR density seen in the ``Lilly-Madau'' diagram. The
ridge evolves independently of galaxy stellar mass to a particular
turnover mass at the high mass end. Galaxies above the turnover mass
show a sharp decrease in SSFR compared to the average at that epoch,
and the turnover mass increases with redshift. 
\end{abstract}


\section*{Introduction}
In an effort to study galaxy assembly, we look at the contribution of
star formation (SF) to the growth of stellar mass in galaxies as a function
of time.  Stellar mass functions reveal that 50\% of the local stellar
mass density was in place by $z=1$ (Drory et al. 2005).  The stellar
mass function describes the build up of stellar mass over cosmic time,
but does not identify possible causes of growth for individual
galaxies: mergers, tidal interactions, internal star formation, etc.

The well known ``Lilly-Madau'' diagram shows the star frormation rate
(SFR) density decreasing by a factor of 10 since $z=1$ (Madau et
al. 1996, Lilly et al. 1996), a result similarly concluded by
measurements from SFR indicators covering nearly the full
electromagnetic spectrum (e.g. Perez-Gonzalez et al. 2005).  These
results, however, give no indication of where the SF is taking place.
Do all galaxies experience a factor of ten decrease in SFR?  Is this
trend dominated by high mass galaxies ending SF or by low mass
galaxies initiating larges amount of efficient SF?  How and where are
half of the local stars formed while the global SFR in decreasing by a
factor of 10?

Cowie et al. (1996) used rest-frame $K$-band (2.2$\mu$m) luminosities
and [OII] $\lambda3727$ equivalent widths to show that galaxies with
rapid SF decrease in $K$ luminosity, and therefore mass, with
decreasing redshift.  A more direct measure of this association is the
the specific star formation rate (SSFR, Guzman et al. 1997, Brinchmann
\& Ellis 2000), which measures the SFR per unit galaxy stellar mass,
to study explicitly how SF contributes to galaxy growth for galaxies
of different masses at different times in the history of the universe.

We combine two complementary redshift surveys to broaden the mass and
redshift range that we can probe.  The wide-area, medium deep MUNICS
(Drory et al. 2001, Feulner et al. 2003) spectroscopic dataset
provides intermediate to high mass galaxies typically in the mass
range of $M_* \geq 10^{10}\Msun$.  The FORS Deep Field (Heidt et
al. 2003, Noll et al. 2004) covers a small portion of the sky very
deeply, contributing $M_*~<~10^{10}$~\Msun~\ galaxies to the sample.

We adopt an $\Omega_M = 0.3$, $\Omega_{\Lambda} = 0.7$, $H_0 = 72\
\mathrm{km\ s^{-1}\Mpc^{-1}}$ cosmology.

\section*{The Galaxy Data}

The galaxies used in this study are gathered from the MUnich
Near-Infrared Cluster Survey (MUNICS; Drory et al. 2001; Feulner et
al. 2003) and the FORS Deep Field (FDF; Heidt et al. 2003; Noll et
al. 2004).  The MUNICS project is a wide-area, medium-deep,
photometric and spectroscopic survey selected in the $K$-band,
reaching $K\sim19.5$, and including $BVRIJK$.  Spectroscopy is complete
to $K\sim17.5$ over 0.25 square degrees and reaches $K=19.5$ for 100
square arcmins.  The spectra cover a wide wavelength range of
$4000-8500$ $\AA$ at $13.2$ $\AA$ (FWHM) resolution,
sampling galaxies in the redshift range of $0.07<z<1$.  Our MUNICS
sample contains 202 objects, which are mostly massive ($M_*>
10^{10}$\Msun) field galaxies with detectable [OII]$\lambda$3727
emission.

The FORS Deep Field (FDF) spectroscopic survey provides low-resolution
spectra with detectable [OII]$\lambda3727$ in the spectral window
($3300-10000$ {$\AA$ at $23$ $\AA$ (FWHM) resolution) to
$z=1.5$.  The FDF survey is $I$-band selected reaching $I_{AB} = 26.8$
with spectroscopy to $I_{AB}=24$.  The FDF covers $7'$~x~$7'$ in eight
bands: $UBgRIzJK$.  Our FDF sample includes 152 galaxies with
detectable SF, and masses of mostly $M_*< 10^{10}$\Msun.

The trends seen among the FDF and MUNICS galaxy surveys exhibit
similar evolution in their overlapping mass range around
$M_*~=~10^{10}$\Msun\ and are therefore suitable to analyze
simultaneously to cover such a large range of galaxy stellar mass.

We developed an automated spectral measurement routine to consistently
determine the fluxes of emission lines.  We use the flux of the
[OII]~$\lambda3727$ emission feature, which remains in the spectral
window to $z=1.5$, as a SFR indicator.  We use the Kennicutt (1998,
Equation 3) conversion from [OII] line luminosity to SFR in units of
solar masses per year.

Stellar masses are determined for each galaxy by fitting a grid of
composite stellar population models of varying age, star formation
history, and dust extinction to multi-wavelength photometry to
determine individual mass-to-light ($M/L$) ratios.  This process is
described in detail in Drory et al. (2004a).

\section*{Specific Star Formation Rates}

Redshift and SFR distributions for both samples can be found in
Figure~1 of Bauer et al. (2005).  The majority of galaxies in the full
samples (52\% for MUNICS and 80\% for FDF) have detectable SF, via the
[OII]$\lambda$3727 emission feature, and the closely matching redshift
distributions indicating little redshift bias in sample selection.
The maximum SFR increases with redshift, and is consistent between the
two samples.

\begin{figure}[ht]
\vskip.2in        
\includegraphics[angle=-90,scale=0.7]{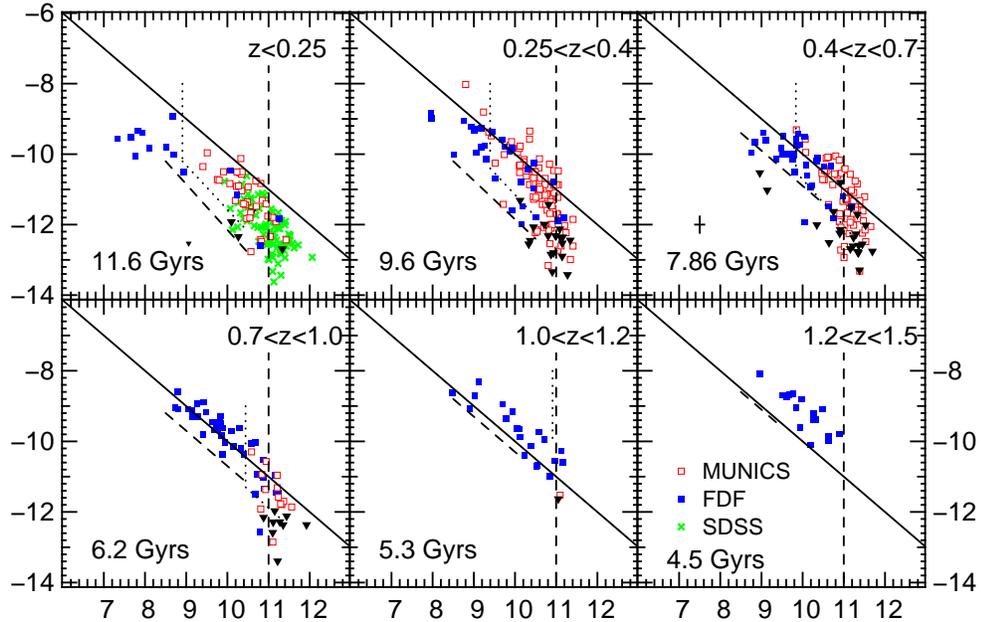}
\caption{Specific Star Formation Rates versus Galaxy Stellar Mass.
The open squares are MUNICS galaxies, filled squares represent FDF
galaxies, the crosses show SDSS galaxies chosen to match the
photometric limits of the MUNICS sample, and the triangles show 1
$\sigma$ upper limits for SSFR.  The solid diagonal line identifies a
constant SFR = 1 \Msun yr$^{-1}$.  The dotted lines show the mass
limits (vertical) and detectable star formation limits for the MUNICS
survey while the dashed lines show the lower limits for detectable
SSFR for FDF, derived from the spectroscopic sensitivity limits.
Estimated error bars, approriate for the majority of galaxies are
shown in the $0.4<z<0.7$ bin.}
\end{figure}

Fig.~1 shows the SSFR versus galaxy stellar mass as a function of
redshift, to $z=1.5$.  The open squares show MUNICS galaxies, the
closed squares are FDF galaxies and the crosses in the lowest redshift
bin are SDSS galaxies to be discussed in more detail later.  Galaxies
of low to intermediate stellar mass form a ``ridge'' in SSFR running
parallel to lines of constant SFR in each redshift bin.  The ridge
exists in all redshift bins until, at some high ``turnover'' mass, the
SSFRs drastically decrease, as expected for the higher mass,
early-type, redder galaxies which tend to form fewer stars after
$z=1$.  The turnover mass increases with redshift, showing higher SSFR
for higher mass galaxies at earlier times, a process termed
``downsizing'' by Cowie et al. (1996).

A correlation exists between SSFR and mass (e.g. Brinchmann \& Ellis
2000).  Here we show this correlation evolves with redshift up to
$z=1.5$ independently of galaxy stellar mass.  The ridge in the SSFR
shifts downward as redshift decreases, indicating a steady decrease in
the global SFR, from 10 $\Msun$yr$^{-1}$ at $z=1.5$ to 1
$\Msun$yr$^{-1}$ at $z=0$, agreeing with concurring trends in the
``Lilly-Madau'' diagram (Madau et al. 1996, Lilly et al. 1996)
determined from a large variety of SFR indicators (see compilation by
Hopkins 2004).  Similar trends are also identified in the large sample
of Feulner et al. (2005).  While detecting the lowest SFRs at any
epoch is affected by incompleteness, the evolution of the upper
envelop is independent of selection effects.

In the present study we apply no corrections for extinction to the
[OII] emission line fluxes.  For the 10\% of the galaxies where we
have the necessary Balmer emission lines to perform a proper dust
correction, we find an average $E(B-V) = 0.2$, assuming a case B
recombination.  We feel it necessary to apply a unique dust correction
to each galaxy (Maier et al. 2005) and consider a flat correction
determined from a small sample of low redshift galaxies irrelevant to
the differential SF studied here.  

To ensure that the increase in SSFR at higher redshifts is not due to
rarer objects being seen as larger volumes are probed, we investigate
the trends in SSFR from the SDSS.  We collected SDSS galaxies
(Abazajian et al. 2004) selected to match the MUNICS magnitude limits
as described in (Drory et al. 2004b), and determined stellar masses in
the same way as the rest of our sample.  We used the reported
[OII]$\lambda$3727 emission line flux from the SDSS and followed the
same Kennicutt (1998) conversion to SFR.

We chose a random sample of SDSS galaxies to match the number of
galaxies in the lowest redshift bin from MUNICS and FDF and show them
in Fig.~1 as crosses.  The consistent SSFR between all galaxies
samples, confirmed by choosing several sets of SDSS galaxies, shows
that our results are not due to volume effects in the low redshift
bin.

\section*{Discussion}

We present a study of the contribution of star formation (SF) to the
growth of stellar mass in galaxies since $z=1.5$ showing the SSFR
versus galaxy stellar mass as a function of redshift over five decades
in galaxy stellar mass.  At all redshifts, the SSFR decreases as
stellar mass increases.  This indicates a higher contribution of SF to
the growth of low mass galaxies since $z=1.5$ and suggests that high
mass galaxies formed the bulk of their stellar content earlier than
$z=1$.

Fig.~1 shows evidence of a ridge in SSFR that runs parallel to lines
of constant SFR.  The ridge exists for all galaxy stellar masses
$M_*=10^{7} - 10^{11}$\Msun\ and increases uniformly, independent of
mass as redshift increases.  The first evidence for such a ridge and
an upper bound in SSFR, moving to higher SFRs with increasing redshift
was noted by Brinchmann \& Ellis (2000).  Our work moves beyond that
study with spectroscopy of a mass limited sample at each redshift bin,
covering a wide range of masses (Bauer et al. 2005).

Galaxies exist on Fig.~1 only when the SF induces detectable amounts
of [OII] $\lambda$3727 emission.  Our mass-selected galaxy samples,
while not biased towards identifying star-forming galaxies, still show
detectable SF in a majority ($50-80\%$) of the galaxies at any epoch.
Many studies focus on disentangling different stages in the lifetimes
of individual galaxies.  While this remains a difficult task, common
trends in evolutionary paths have identified stages where detectable
star formation occurs.  One such stage is represented by a population
of highly star forming, dust enshrouded, luminous IR galaxies (LIRGs)
which are known to be much more common at $z>1$ than they are today
(e.g. Flores et al. 1999, Perez-Gonzalez et al. 2005).  It is most
likely that LIRGs correspond to short bursts of intense SF induced by
recent merging or gas infall (Hammer et al. 2005).  These brief bursts
represent only occasional phases and not normal stages of SF in galaxy
lifetimes, and we could possibly be detecting the low mass galaxies
during LIRG phases.  

The first studies of the evolution of the SFR from \textit{Spitzer
Space Telescope} 24~$\mu$m data have recently been published (Bell et
al. 2005, Perez-Gonzalez et al. 2005).  The SSFR evolution with
stellar mass and redshift largely agrees with the work presented here
for the low to intermediate mass galaxies, but show more scatter
towards higher SSFRs in the high mass regime.  While thermal IR
observations detect the highly dust obscured objects missed by
optical- and NIR-selected samples, a large uncertainty remains in
converting the observed light into the total IR flux.  The amount of
dust heated by old stars is unknown.  Since the most massive galaxies
contain the largest populations of old stars locally (e.g. Kennicutt
et al. 1994) and at high redshifts (Drory et al. 2005), there is a
possible tendency to overestimate the IR-derived SFR at higher stellar
masses.

Large new surveys greatly improve our working understanding of galaxy
evolution and stellar mass build up in the universe, but many
uncertainties persist while we seek to understand the contributing
components of new data.  It remains important to continue
multiwavelength studies of galaxy evolution in seeking concordance
among various methodologies.

\end{document}